\documentclass{ws-ijmpe}
\usepackage{amssymb}
\usepackage{braket}
\usepackage[super,compress]{cite}
\usepackage{epsfig}
\usepackage{epstopdf}
\begin{document}


\catchline{}{}{}{}{}

\title{\textbf{Study of Gamow-Teller strength and associated weak-rates on odd-A nuclei in
stellar matter}}

\author{Muhammad Majid\footnote{Corresponding author}, Jameel-Un Nabi and Muhammad Riaz}
\address{Faculty of Engineering Sciences,\\GIK
Institute of Engineering Sciences and Technology, Topi 23640, Khyber
Pakhtunkhwa, Pakistan\\ Email: {\em $^{*}$majid.phys@gmail.com,
jameel@giki.edu.pk} }

\maketitle

\begin{history}
\received{Day Month Year}
\revised{Day Month Year}
\end{history}

\begin{abstract}
In a recent study by Cole et al. \cite{Col12}, it was concluded that
QRPA calculations show larger deviations and overestimate the total
experimental Gamow-Teller (GT) strength. It was also concluded that
QRPA calculated electron capture rates exhibit larger deviation than
those derived from the measured GT strength distributions. The main
purpose of this study is to probe the findings of the Cole et al.
paper. This study gives useful information on the performance of
QRPA-based nuclear models. As per simulation results, the capturing
of electrons that occur on medium heavy isotopes have a significant
role in decreasing the ratio of electron-to-baryon content of the
stellar interior during the late stages of core evolution. We report
the calculation of allowed charge-changing transitions strength for
odd-A $fp$-shell nuclei ($^{45}$Sc and $^{55}$Mn) by employing the
deformed pn-QRPA approach. The computed GT transition strength is
compared with previous theoretical calculations and measured data.
For stellar applications the corresponding electron capture rates
are computed and compared with rates using previously calculated and
measured Gamow-Teller values. Our finding show that our calculated
results are in decent accordance with measured data. At higher
stellar temperature our calculated electron capture rates are larger
than those calculated by Independent Particle Model (IPM) and shell
model. It was further concluded that at low temperature and high
density regions the positron emission weak-rates from $^{45}$Sc and
$^{55}$Mn may be neglected in simulation codes.
\end{abstract}

\keywords{Gamow-Teller transitions; electron capture rates; pn-QRPA
theory; stellar evolution; core-collapse}

\ccode{PACS numbers: 23.40.Bw, 26.30.Jk, 23.40.-s, 97.10.Cv,
26.50.+x}


\section{Introduction}
Study of supernova process is one of the means to explore our
Universe. All type of natural interactions manifest themselves in a
supernova explosion.  The study of these interactions explain many
scenarios of the universe \cite{Sar13}. The strong and the weak
interactions, during hydrostatic process of stellar evolution,
produce nuclei till mass number A$\approx 60$. More massive nuclei
are most probably synthesized in supernova explosion environments at
very high temperature and density. To date the complete mechanism of
supernova explosion is not well understood. There are many
complexities involved. Researchers world-wide continue their quest
for a deeper understanding of the dynamics of core-collapse.

The weak interaction processes are considered as to be the key
parameters for understanding the mechanism of late stages of stellar
evolution \cite{Ffn80}. These processes also perform a vital role to
estimate the pre-supernova stellar core composition as well as the
nucleosynthesis of neutron-rich (heavier) nuclei \cite{Ffn80,
Bur57}. The Gamow-Teller (GT) charge-changing transitions are
responsible for these processes, due to which the GT distribution of
$fp$-shell nuclei are of particular significance. In pre-supernova
structure these nuclei are the major ingredient of the astrophysical
core leading to thermonuclear/type-Ia or core-collapse/type-II
supernovae \cite{Auf94, Heg01}. In the pre-explosion development of
core-collapse supernovae,  once the electrons Fermi energy
sufficiently raised to overcome Q-value limitations that confine
electron capture (EC) under laboratory environment, the nuclear
content in the astrophysical core is neutronized and EC reactions
reduce the lepton-to-baryon fraction (Y$_e$)\cite{bethe79}. In other
words the EC process reduces the electron degeneracy pressure,
electron-to-baryon  ratio and entropy of the system. During these
stages of stellar evolution, EC on $fp$-shell nuclei play important
role, since these are the key ingredients of the stellar core in
pre-supernovae formation which lead to core-collapse  or
thermonuclear supernovae.\\
During later phases of stellar evolution, electrons have enough
energy to initiate transitions to GT resonance. For probing GT
transitions at higher  excitation  energy  (n, p), (p, n),($^3$He,
t) and (d, $^2$He) reactions are studied.  Williams et al.
\cite{wil95} observed that the total charge-changing GT strength in
the EC direction was quenched (compared to theoretical estimates)
and distributed over many final levels in the daughter nuclei due to
the residual nucleon-nucleon correlation. More efficient way for
extraction of GT$_+$ (responsible for EC process) strength in stable
nuclei is to investigate charge-exchange reactions \cite{ost92}. In
case of incident energy higher than 100 MeV, the dominating
components of effective interaction are iso-vector spin-flip and
spin-isospin transitions. The momentum transfer in forward angle is
rather small and GT operator with $\Delta T=1$, $\Delta J^+ = 1^+$,
$\Delta L = 0$ in reaction cross section may dominate. The cross
sections, in region of zero momentum transfer, are directly related
to strength of $\beta$-decay between the same states. In order to
study GT strength distributions, charge-changing reactions at low
momentum transfer is to be employed in case where $\beta$-decay is
not possible.\\
Researchers are working hard to estimate, in a microscopic fashion,
the ground- and excited-state charge-changing transition strength
distributions. Measurement of GT functions itself is a challenging
task. Due to involvement of hundreds of nuclei in stellar matter and
also in order to incorporate finite-temperature effects, theoretical
predictions of GT distributions continue to be the affordable
choice. Because of the significant implications of the weak rates in
astrophysical scenario, they were widely studied using various
nuclear models. The first considerable attempt to measure the
astrophysical weak rates over a wide range of temperature and
density was done by Fuller, Fowler and Newman (FFN)~\cite{Ffn80}.
The calculation was based on the independent particle model and the
authors employed the Brink-Axel hypothesis \cite{Bri65} in their
calculation to estimate the excited state GT strength functions.
They further incorporated measured data available at that time in
their calculation to improve the reliability of the results. Later,
the FFN work was expanded for heavier nuclei with $A
> 60$ by Aufderheide and collaborators~\cite{Auf94}. Shell model
Monte Carlo (SMMC) approach was employed for the first time to
calculate, in a fully microscopic way, the Gamow-Teller
contributions to EC rates for \emph{fp}-shell nuclei, by considering
the thermal effects \cite{who}. The electroweak interaction matrix
elements were calculated in the limit of zero-momentum transfer,
with the GT operators as the main constituent. However, the shell
model diagonalization cannot be performed beyond the \emph{fp}-shell
nuclei, because of the huge dimension of the model space involved.
To overcome this limitation a hybrid model (SMMC + RPA) was
introduced \cite{Lan001, Sam03}. Dzhioev et al., introduced an
alternative approach known as thermal quasi-particle random-phase
approximation (TQRPA) model, for the calculation of electron capture
rates of hot nuclides. This is basically a statistical approach to
the nuclear many-body problem at finite-temperature \cite{Dzh09,
Dzh10}. In TQRPA model, instead of calculating the individual
strength functions for the nuclear ground and excited states, one
calculates an 'average' temperature dependent strength distribution.
A TQRPA model, based on the Woods-Saxon potential and separable
multi-pole and having spin-multi-pole particle-hole interactions,
was used for the thermal evolution of GT$_{+}$ distributions on
$^{54,56}$Fe and $^{76,78,80}$Ge nuclei \cite{who}. This model was
based on the thermofield dynamics (TFD) formalism and performed
calculation of the weak-interaction response of nuclei at finite
temperature. It was concluded that thermal effects shift the
GT$_{+}$ centroid to lower excitation energies \cite{Dzh10}.
Recently, a self-consistent finite-temperature RPA (FTRPA) model
based on Skyrme functionals has been applied in Refs. \cite{Paa09,
Fan12} to study EC cross sections and weak rates using several
different interactions. Furthermore, a similar approach, extended to
the relativistic framework i.e., finite-temperature RPA (FTRRPA)
model has been employed in Ref. \cite{Niu11}. It was concluded that
the FTRRPA provides a universal theoretical tool for the analysis of
stellar weak-interaction processes in a fully consistent microscopic
approach, specifically for region of neutron-rich nuclei. In this
work we have used the pn-QRPA model having separable residual GT
interactions in a deformed basis, to compute the allowed GT
transitions and stellar weak reaction rates for the odd-A nuclei
$^{45}$Sc and $^{55}$Mn. In our model we calculate the ground and
excited state temperature dependent nuclear partition functions. We
compute ground and excited states of parent and daughter nuclei and
the nuclear matrix elements connecting these states via the GT
operator within the QRPA approach. We assume that the probability of
excited states follow a normal Boltzmann distribution. However,
currently we are unable to determine the complete finite-temperature
effects on the GT transition functions (applicable at high stellar
temperatures exceeding 10$^{10}$ K \cite{Rau00}) and on the pairing
correlations. In our model the pairing gap is considered to be
independent of temperature. In our model the nuclear Fermi surface
is smeared due to pairing correlations only. We cite this as a
short-coming of our current model. We hope to include the
finite-temperature effects in our model as a future assignment. It
should be mentioned that the thermal QRPA approaches and our model
do not rely
on the Brink's hypothesis, as used in shell model calculations.\\
Cole et al. \cite{Col12} have presented a systematic evaluation of
the capability of theoretical nuclear models to reproduce the
measured GT strength of charge-exchange reactions at intermediate
energies. The authors have concluded that the GT strength
distributions calculated in the shell models reproduce well the
measured data, whereas QRPA calculations \cite{Mol90} show larger
deviations and overestimate the total experimental GT strength. It
was also concluded that EC rates from the shell-model calculations
are also much closer to the EC rates derived from the experimental
GT strength distributions than those calculated on basis of the QRPA
framework. The current study probes the conclusion of the Cole et
al. study and provide useful information on the performance of
QRPA-based models and refines the conclusions in Ref. \cite{Col12}.
Our findings show that this is not the case for all kind QRPA
calculations and form the prime motivation for this paper. In the
next section we briefly describe the necessary formalism of pn-QRPA
model and its parameters used for the computation of transitions
strength and stellar rates. Sec.~3 presents the comparison of our
calculation with different experimental and theoretical results.
Finally the conclusion of our findings is describe in Sec.~4.

\section{Theoretical Formalism}

The pn-QRPA Hamiltonian is specified by
\begin{equation} \label{GrindEQ__1_}
H^{pn-QRPA} =H^{sp} + V^{pairing} + V_{GT}^{ph} + V_{GT}^{pp} ,
\end{equation}
where $H^{sp}$ is the single particle Hamiltonian, $V^{pairing}$ is
pairing force (considered within the BCS approximation),
$V_{GT}^{ph}$ and $V_{GT}^{pp}$ represent the particle-hole ($ph$)
and particle-particle ($pp$) interaction parameters for GT channel,
respectively. The single particle eigenfunctions and eigenvalues
were computed in Nilsson model \cite{Nil55}, in which the nuclear
deformation parameter ($\beta_{2}$) was incorporated. The $ph$ and
$pp$ interaction strengths were characterized by model parameters
$\chi$ and $\kappa$, accordingly. These parameters were selected
with the constraint that the computed Gamow-Teller strength satisfy
the model independent Ikeda sum rule (ISR) \cite{Isr63}. Other
parameters needed for computation of weak-interaction rates are the
pairing gap ($\Delta _{nucleon}$), the Nilsson potential parameters
(NPP) and the Q-values. The NPP were chosen from \cite{Rag84} and
$\hbar \omega = 41/A^{1/3}$ in units of MeV was considered for
Nilsson oscillator constant, similar for neutrons and protons. The
calculated half-life (T$_{1/2}$) values rely weakly on the pairing
gaps ($\Delta _{nucleon}$) between nucleons \cite{Hir91}. The same
were calculated using
\begin{equation}
\Delta _{n} =\Delta _{p} =12/\sqrt{A}  (MeV),
\end{equation}
$\beta_{2}$ was determined by using the formula
\begin{equation}
\beta_{2} = \frac {125 (Q_{2})} {1.44 (A)^{2/3} (Z)},
\end{equation}
where $Q_{2}$ denote the electric quadrupole moment taken from Ref.
\cite{Mol16}. Q-values of reactions were taken from Ref.
\cite{Aud12}.

In our model the charge-changing transitions are described by phonon
creation operators defined by
\begin{equation}
A_{\omega}^{+}(\mu)=\sum_{pn}(X^{pn}_{\omega}(\mu)a_{p}^{+}a_{\bar{n}}^{+}-Y_{\omega}^{pn}(\mu)a_{n}
a_{\bar{p}}).
\end{equation}
The summation is taken over all the p-n pairs having $\mu$ =
\textit{m$_{p}$-m$_{n}$} = 0, $\pm$1, here
\textit{m$_{n}$}(\textit{m$_{p}$}) depicts the angular momentum
third component of the for neutron(proton). The
\textit{a$^{+}_{n(p)}$} are the creation operator of a
quasi-particle (q.p) state of neutron(proton). The
\textit{$\bar{p}$} (\textit{$\bar{n}$}) represents the time reversed
state of \textit{p} (\textit{n}). The ground level of our model with
respect to the QRPA phonons is considered as the vacuum,
A$_{\omega}(\mu)|QRPA\rangle$ = 0. The excitation energy ($\omega$)
and amplitudes (\textit{X$_{\omega}, Y_{\omega}$}) of phonon
operator were obtained by solving the  RPA equation. Detailed
solution of RPA matrix equation may be seen in Refs. \cite{Hir91,
Mut89}.
\subsection{Quasi-particle transitions}
The RPA is formulated for excitations from the J$^{\pi}$ = 0 ground
level of an even-even nucleons. When the parent nuclide consists of
odd nucleon, as in the current work, then the ground level is
represented as a one-quasi-particle state, in which the odd
quasi-particle occupies the single quasi-particle orbit of the
smallest energy. Two kinds of transitions are possible in our model.
One is the phonon excitations alone, in which the quasi-particle
acts merely as a spectator. The other is transitions of the
quasi-particle itself, and phonon correlations to the quasi-particle
transitions in first order perturbation were introduced using the
Refs. \cite{Hal67, Ran73}. The phonon-correlated one-quasi-particle
states are specified by:
\begin{equation}
\begin{split}
\ket{p_{corr}}=a^+_p\ket{-}+\sum\limits_{n,\omega}a^+_nA^+_{\omega}(\mu)\ket{-}\bra{-}[a^+_nA^+_{\omega}(\mu)]^+H_{qp-ph}a^+_p\ket{-}E_p(n,\omega)\\
\ket{n_{corr}}=a^+_n\ket{-}+\sum\limits_{p,\omega}a^+_pA^+_{\omega}(-\mu)\ket{-}\bra{-}[a^+_pA^+_{\omega}(-\mu)]^+H_{qp-ph}a^+_n\ket{-}E_n(p,\omega),
\end{split}
\label{npcorr}
\end{equation}
with
\begin{equation}
E_{a}(b,\omega)=\dfrac{1}{\epsilon_{a}-\epsilon_{b}-\omega}
\label{Ea}
\end{equation}
Eq.~\ref{npcorr} has two parts, first part of the equation
represents the proton (neutron) quasi-particle state and the second
part shows the admixture of correlation of RPA phonon by
quasi-particle phonon coupling Hamiltonian (H$_{qp-ph}$), obtained
from the separable $pp$ and $ph$ forces by the Bogoliubov
transformation \cite{Mut89}. The sum in Eq.~5 run over all the
levels of phonon and proton (neutron) quasi-particle levels that
fulfill the $\pi_p.\pi_n=1$ and $m_p-m_n=\mu$. The analytical
treatment of the quasi-particle transition amplitudes for correlated
states (for the general force and charge changing transitions mode)
may be seen in Refs. \cite{Mut89, Sta90, Mut92}.

In case of odd-A nucleus, low- lying states were obtained by lifting
the quasi-particle in the orbit of the smallest energy to
higher-lying orbits. Parent states of $^{45}$Sc and $^{55}$Mn
(even-neutron and odd-proton system) were represented by
three-proton states or one-proton two-neutron states, corresponding
to excitation of a neutron or proton
\begin{equation}
\begin{split}
\ket{p_1p_2p_{3corr}}=a^+_{p_1}a^+_{p_2}a^+_{p_3}\ket{-}+\dfrac{1}{2}\sum\limits_{p'_1p'_2n'\omega}a^+_{p'_1}a^+_{p'_2}a^+_{n'}A^+_{\omega}(\mu)\ket{-}\\
\bra{-}[a^+_{p'_1}a^+_{p'_2}a^+_{n'}A^+_{\omega}(\mu)]^+H_{qp-ph}a^+_{p_1}a^+_{p_2}a^+_{p3}\ket{-}E_{p_1p_2p_3}(p'_1p'_2n',\omega)
\end{split}
\label{p1p2p3corr}
\end{equation}
\begin{equation}
\begin{split}
\ket{p_1n_1n_{2corr}}=a^+_{p_1}a^+_{n_1}a^+_{n_2}\ket{-}+\dfrac{1}{2}\sum\limits_{p'_1p'_2n'\omega}a^+_{p'_1}a^+_{p'_2}a^+_{n'}A^+_{\omega}(-\mu)\ket{-}\\
\bra{-}[a^+_{p'_1}a^+_{p'_2}a^+_{n'}A^+_{\omega}(-\mu)]^+H_{qp-ph}a^+_{p_1}a^+_{n_1}a^+_{n_2}\ket{-}E_{p_1n_1n_1}(p'_1p'_2n',\omega)+\dfrac{1}{6}\\
\sum\limits_{n'_1n'_2n'_3\omega}a^+_{n'_1}a^+_{n'_2}a^+_{n'_3}A^+_{\omega}(\mu)\ket{-}\bra{-}[a^+_{n'_1}a^+_{n'_2}a^+_{n'_3}A^+_{\omega}(\mu)]^+\\
H_{qp-ph}a^+_{p_1}a^+_{n_1}a^+_{n_2}\ket{-}E_{p_1n_1n_2}(n'_1n'_2n'_3,\omega),
\end{split}
\label{p1n1n2corr}
\end{equation}
with the energy denominators of first order perturbation
\begin{equation}
E_{abc}(def,\omega)=\dfrac{1}{(\epsilon_a+\epsilon_b+\epsilon_c)-(\epsilon_d+\epsilon_e+\epsilon_f+\omega)}
\end{equation}
The three quasi-particle states to odd neutron and even proton
nuclide are obtained from (Eq.~\ref{p1p2p3corr} and
Eq.~\ref{p1n1n2corr}) by the interchange of proton states and
neutron states, $p\longleftrightarrow n$ and
$A^+_{\omega}(\mu)\longleftrightarrow A^+_{\omega}(-\mu)$.
The excited states for an even-proton and odd-neutron nuclei
(daughter nuclei) were
constructed as\\
(1) raising odd neutron from the lower (ground)state to the upper
(excited) state (one quasi-particle states state),\\
(2) three-neutron states, corresponding to excitation of a
neutron or,\\
(3) two-proton and one-neutron states, corresponds to
excitation of proton.\\
The formulae for multi-quasi-particle GT transitions and their
reduction to correlated (c) one-quasi-particle states were defined
as:
\begin{equation}
\begin{split}
\bra{p^f_1n^f_1n^f_{2c}}t_{\pm}\sigma_{\mu}\ket{n^i_1n^i_2n^i_{3c}}=\delta(n^f_1,n^i_2)\delta(n^f_2,n^i_3)\bra{p^f_{1c}}t_{\pm}\sigma_{\mu}\ket{n^i_{1c}}\\
-\delta(n^f_1,n^i_1)\delta(n^f_2,n^i_3)\bra{p^f_{1c}}t_{\pm}\sigma_{\mu}\ket{n^i_{2c}}\\
+\delta(n^f_1,n^i_1)\delta(n^f_2,n^i_2)\bra{p^f_{1c}}t_{\pm}\sigma_{\mu}\ket{3^i_{3c}}
\end{split}
\end{equation}
\begin{equation}
\begin{split}
\bra{p^f_1n^f_1n^f_{2c}}t_{\pm}\sigma_{-\mu}\ket{p^i_1p^i_2n^i_{1c}}=\delta(p^f_1,p^i_2)[\delta(n^f_1,n^i_1)\bra{n^f_{2c}}t_{\pm}\sigma_{-\mu}\ket{p^i_{1c}}\\
-\delta(n^f_2,n^i_1)\bra{n^f_{1c}}t_{\pm}\sigma_{-\mu}\ket{p^i_{1c}}]\\
-\delta(p^f_1,p^i_1)[\delta(n^f_1,n^i_1)\bra{n^f_{2c}}t_{\pm}\sigma_{-\mu}\ket{p^i_{2c}}\\
-\delta(n^f_2,n^i_1)\bra{n^f_{1c}}t_{\pm}\sigma_{-\mu}\ket{p^i_{2c}}]
\end{split}
\end{equation}
\begin{equation}
\begin{split}
\bra{p^f_1p^f_2p^f_{3c}}t_{\pm}\sigma_{\mu}\ket{p^i_1p^i_2n^i_{1c}}=\delta(p^f_2,p^i_1)\delta(p^f_3,p^i_2)\bra{p^f_{1c}}t_{\pm}\sigma_{\mu}\ket{n^i_{1c}}\\
-\delta(p^f_1,p^i_1)\delta(p^f_3,p^i_2)\bra{p^f_{2c}}t_{\pm}\sigma_{\mu}\ket{n^i_{1c}}\\
+\delta(p^f_1,p^i_1)\delta(p^f_2,p^i_2)\bra{p^f_{3c}}t_{\pm}\sigma_{\mu}\ket{n^i_{1c}}.
\end{split}
\end{equation}
Here $\overrightarrow{\sigma}$ and $\tau_{\pm }$ denote the spin and
the isospin operators, respectively. For further necessary details
Ref. \cite{Mut89} may be seen.
\subsection{Weak-decay rates} The positron emission (PE) and
electron capture (EC) weak-rates from the \emph{nth} parent state to
\emph{mth} daughter state is given by
\begin{equation}
 \lambda _{nm}^{EC (PE)} =\ln 2\frac{f_{nm}^{EC (PE)}(T,\rho
 ,E_{f})}{(ft)_{nm}},
\end{equation}
the term $(ft)_{nm}$ is linked to the reduced transition probability
($B_{nm}$) by
\begin{equation}
(ft)_{nm} =D/B_{nm}.
\end{equation}
The reduced transition probabilities $B_{nm}$'s are sum of GT and
Fermi transition probabilities and given by
\begin{equation}
 B_{nm}=((g_{A}/g_{V})^{2} B(GT)_{nm}) + B(F)_{nm}.
\end{equation}
The value of constant $D$ was taken as 6143 $s$ Ref. \cite{Har09}
and $g_{A}/g_{V}$ was taken as -1.2694. The reduced Fermi and GT
transition probabilities were expressed as
\begin{equation}
B(F)_{nm} = \frac{1}{2J_{n} +1} \langle{m}\parallel\sum\limits_{k}
\tau_{\pm}^{k}\parallel {n}\rangle|^{2}
\end{equation}
\begin{equation}
B(GT)_{nm} = \frac{1}{2J_{n} +1} \langle{m}\parallel\sum\limits_{k}
\tau_{\pm}^{k}\overrightarrow{\sigma}^{k}\parallel {n}\rangle|^{2}.
\end{equation}
For the construction of daughter and parent excited levels and
computation of GT nuclear matrix elements we refer to \cite{Nab99}.

The phase space ($f_{nm}$) integrals (over total energy) were
calculated (by adopting the natural units $\hbar=m_{e}=c=1$) as
\begin{equation}
f_{nm}^{EC} = \int _{w_{1} }^{\infty }w\sqrt{w^{2} -1} (w_{m}
+w)^{2} F(+ z, w) R_{-}dw,
\end{equation}\begin{equation}\label{ps}
f_{ij}^{PE} = \int_{1}^{w_{m}} w \sqrt{w^{2}-1} (w_{m}-w)^{2} F(-
Z,w) (1-R_+) dw,
\end{equation}

In above equations $w$ is the (K.E + rest mass) of electron and
threshold energy of EC is represented by $w_{l}$. R$_{\pm}$ shows
the distribution function of positrons (electrons) and given by
\begin{equation}
 R_{-} =\left[\exp \left(\frac{E-E_{f} }{kT} \right)+1\right]^{-1}
\end{equation}
\begin{equation}
R_{+} =\left[\exp \left(\frac{E+2+E_{f} }{kT} \right)+1\right]^{-1},
\end{equation}
where $E_{f}$, $E (= (w - 1))$, and T denote the Fermi energy,
kinetic energy of electrons, and temperature, respectively. The
Fermi functions denoted by $F (+Z, w)$ were computed using the
recipe of Ref. \cite{Gov71}. If the positron (or electron) emission
total energy ($w_{m}$) value was larger than -1, then $w_{l}$ was
taken as 1, and if $w_{m}$ $\leq$ 1, then $w_{m}=|w_{l}|$. $w_{m}$
is given by
\begin{equation} w_{m} = m_{p} -m_{d} + E_{n} -E_{m},
\end{equation}
\noindent where $m_{p}$ and ($m_{d}$) are the masses of parent and
daughter nuclei, respectively and their corresponding excitation
energies are denoted by $E_{n}$ ($E_{m}$), respectively. The
interior temperature of the stellar core is high enough and there is
always a chance of parent excited levels occupation. The EC and PE
total weak-interaction rates were calculated using
\begin{equation}
\lambda^{EC(PE)} =\sum _{nm}P_{n} \lambda _{nm}^{EC(PE)}.
\label{rates}
\end{equation}
where $P_{n}$ represents the probability of occupation of parent
excited states and follows the normal Boltzmann distribution. In
Eq.~\ref{rates}, the summation was applied over all final and
initial states until satisfactory convergence in EC(PE) rates were
obtained.

\section{Results and Discussion}

Our model was able to calculate GT transitions from 200 discrete
parent levels to 300 daughter levels. Our model calculates GT
transition strength at energy range of (E$\pm$dE) where "E"
represent the excitation energy in daughter nucleus and "dE" is
equal to 0.05 MeV. Within this energy range our model determines all
those GT transitions (which satisfy the allowed selection rules),
adds them and put the cumulative strength corresponding to
excitation energy "E". The charge-changing transitions from the
ground level of $^{45}$Sc (parent nucleus) to $^{45}$Ca (daughter
nucleus) levels is shown in Fig.~1. The transition strength linking
the ground level of $^{55}$Mn to $^{55}$Cr in the GT$_{+}$ direction
is shown in Fig.~2. The comparison of pn-QRPA computed GT
distribution strength with experiment and previous theoretical
calculations are also presented in Figs.~(1~-~2). There are six
panels in each figure. Panel~(a) shows the (n,p) GT data of the
experiment performed by Alford and collaborators \cite{Alf91}. Data
were only presented in energy bins (E$_{ex}$) of 1 MeV due to the
limited resolution of charge-changing (n,p) reaction in the
experiment. It was further noted by the authors in Ref. \cite{Alf91}
that the upper limit of the extracted charge-changing strength in
daughter nuclei ($^{45}$Ca, $^{55}$Cr) was up to E$_{ex}$ of 2 MeV
possibly due to the contamination on hydrogen in the target.
Panel~(b) depicts the GT distribution computed by employing the
current pn-QRPA approach. We used a quenching factor (f$_{q}$) value
of 0.6. The same $f_{q}$ value of 0.6 was suggested for the RPA
results in case of $^{54}$Fe \cite{Vet89,Ron93}. Panel~(c) shows the
QRPA calculation of GT distribution using the Skyrme interactions
\cite{Sar13} (shown as QRPA(S)). Sarriguren used a f$_{q}$ value of
0.7 and the Skyrme force SLy4 \cite{Cha98} for this calculation.
Panel~(d) depicts the theoretical GT transition strength using QRPA
formalism of Ref. \cite{Mol90} (shown as QRPA(M)) employing the
deformation and masses achieved from the finite-range droplet model
(FRDM) \cite{Mol95}. Results of QRPA(M) are divided by a factor of
3. Panels~(e) and (f) show the spin-isospin transitions using
shell-model employing the interactions KB3G \cite{Pov01} and GXPF1a
\cite{Hon04}, accordingly. The shell model results used a f$_{q}$
value of 0.74 in their calculation. It is noted from figure~1 that
our calculated strength distribution is well fragmented and, unlike
previous theoretical estimates, do not put bulk of strength in one
single transition. For the case of $^{55}$Mn, Fig.~2 shows that the
QRPA models calculate bigger GT transitions as compared to shell
model results.

The re-normalized Ikeda sum rule (ISR$_{re}$) \cite{Isr63} is given
by
\begin{equation}
ISR_{re} = \sum B(GT_{-}) - \sum B(GT_{+})\cong 3f_{q}^{2}(N-Z).
\label{Eqt. ISR}
\end{equation}
Table~1 shows that the current pn-QRPA model satisfies the
ISR$_{re}$. The $B(GT_{\pm})$ values appearing in third and fourth
column of Table~1 were calculated using Eq.~17.

The deformed pn-QRPA computed electron capture (EC) and positron
emission (PE)  weak-rates for $^{45}$Sc and $^{55}$Mn, at selected
stellar density and temperature values, are shown in Tables~(2~-~3),
respectively. The weak-rates are presented for stellar temperature
range (0.7--30)$\times 10^{9}$ K at stellar densities ($10^{2}$,
$10^{5}$, $10^{8}$ and $10^{11})$ gcm$^{-3}$. The calculated EC and
PE rates (Eq.~\ref{rates}) are stated in $\log_{10}$ values (in
units of s$^{-1}$). The weak rates increase as the stellar
temperature rises. This increase is due to the fact that the
occupation probability of parent excited states increase with rising
of stellar core temperature. As the core stiffens, the electron
Fermi energy level rises. This leads to sizeable increase of EC
rates at high stellar density. The PE rates remains more or less
constant as the stellar core stiffens. In our calculation it is
assumed that positrons generate via electron-positron pair
creation, at high stellar temperatures ($kT > 1$ MeV). The complete
set of EC and PE rates for $^{45}$Sc and $^{55}$Mn, on a fine
density-temperature grid useful for interpolation processes, may be
demanded as ASCII files from the authors. It is noted that at low
temperatures and high stellar density regions the PE weak-rates are
orders of magnitude smaller than the EC rates and may be safely
omitted in simulation codes.

The comparison of our computed stellar EC rates with
\textit{ground-state} EC reaction rates are displayed in
Figs.~(3~-~4). Fig.~3 shows the result of $^{45}$Sc at stellar
densities (10$^{7}$ and 10$^{9}$) gcm$^{-3}$, while similar results
are depicted  for $^{55}$Mn in Fig.~4. In both figures we show the
pn-QRPA calculated EC rates due to (i) 200 excited state GT
distributions calculated by our pn-QRPA model and \textit{ground
state} (ii) measured GT strength distribution, (iii) QRPA(M)
computed transitions strength distribution, (iv) shell model (SM)
(KB3G interaction) calculated GT strength distribution and (v) SM
(GXPF1a interaction) calculated GT strength distribution. The
corresponding references were stated earlier. Apart from the pn-QRPA
data (where we microscopically calculate excited levels GT strength
distributions), all EC rates were computed using only the ground
state GT strength distributions and were adopted from Ref.
\cite{Col12}. The EC rates are plotted in $\log_{10}$ values (in
units of s$^{-1}$) as a function of stellar temperature (T$_{9}$ =
10$^{9}$ K). On abscissa the range of temperature varies from
(T$_{9}$ =~2--10). These regions of density and stellar temperature
are important for stellar scenarios associated with silicon burning
phases (T$_{9}$ $\sim$ 3, $\rho~Y_{e}$ $\sim$ 10$^{7}$ gcm$^{-3}$),
for phenomena related to type-Ia supernova, up to pre-collapse of
the core (T$_{9}$ $\sim$ 10, $\rho~Y_{e}$ $\sim$ 10$^{9}$
gcm$^{-3}$). For the case of $^{45}$Sc, at (T$_{9}$ = 2, and
$\rho~Y_{e}$ $\sim$ 10$^{7}$ gcm$^{-3}$) our computed weak rates are
in decent comparison with measured data, but an order of magnitude
greater than shell model calculation. However at higher stellar
temperature our calculated weak rates are bigger almost by factor of
5 than the data. The enhancement comes because our calculation also
takes into consideration GT transitions from parent excited states
that show their effect at high T$_{9}$ values. The QRPA(M) rates are
biggest because of calculation of big GT strength distribution (see
Fig.~1). Shell model (SM) weak interaction rates are much smaller.
It was reported in Ref. \cite{Col12} that for the case of $^{45}$Sc,
largest discrepancy was noted between measured data and shell model
results. For the case of $^{55}$Mn the theoretical estimates match
rather well with experimental data (Fig.~4) at $\rho~Y_{e}$ $\sim$
10$^{7}$ gcm$^{-3}$. At low temperature pn-QRPA calculated weak
rates are smaller than the data and shell model calculation almost
by factor of 12 and 7, respectively. This may be attributed to the
fact that within our model, the bulk of GT strength is calculated at
high excitation energies (see Fig. 2). At low stellar temperatures
the calculated EC rates are very sensitive to the strength
distributions at low excitation energies. However at high stellar
temperatures our calculated weak-rates exceed from the EC rates
calculated by using the experimental GT strength and shell model
calculations, by factor of almost 3 and 4, respectively. Because at
higher temperature the low-lying strength becomes less significant
as the capture proceeds mainly due to the GT resonance. Also at
higher temperature the occupation probability of excited states are
significant.

Next we compare the pn-QRPA calculated stellar EC rates with
previous calculations of stellar EC rates. Excited parent states
contribution were taken into consideration by the large scale shell
model (LSSM)\cite{Lan01} and Independent Particle Model (IPM)
\cite{Ffn80} calculations. The IPM calculation was supplemented with
measured transitions strength from $\beta$-decay reaction
experiments. The mutual comparison is shown in Figs.~(5~-~6). The
LSSM weak interaction rates were used in the simulation of
presupernova stellar evolution having masses in the range
11-40\;M$_\odot$ \cite{Heg01}, while IPM results were employed in
several simulation codes (e.g. KEPLER stellar evolution code
\cite{Wea78}). Both figures show three panels. The top panel depicts
the comparison of calculated EC rates at stellar density ($\rho
Y_{e}$) = 10$^{4}$ gcm$^{-3}$ corresponding to low density region.
The middle and bottom panels depict the comparison at stellar
densities 10$^{8}$ gcm$^{-3}$ and 10$^{11}$ gcm$^{-3}$,
respectively. These correspond, respectively, to medium and high
density regions of the core.  In IPM calculation, the unmeasured
matrix elements for GT strength were given an average value of $log
ft$ equal to 5. The results of pn-QRPA and LSSM models are
microscopic in nature and show a more realistic image of the
phenomena occurring in the astrophysical situations. The overall
mutual comparisons show that at low temperatures pn-QRPA computed EC
weak reaction rates are in good agreement with LSSM calculated
weak-rates. However at high temperatures, as the occupation
possibility of excited levels are significant, our calculated EC
weak-rates are bigger. In stellar matter these enhanced EC rates may
have substantial impact during the  late phases of presupernova
evolution of high mass stars. At high temperatures, the LSSM
calculated values are too small as compared to other calculations,
this behavior of LSSM rates can also be seen in Ref.\cite{Maj16}, in
case of chromium isotopes. The Lanczos-based approach was used by
LSSM, which is the main cause for this discrepancy and this was also
pointed by Ref. \cite{Pru03}. The computed weak rates of LSSM are
function of Lanczos iterations, that are required for the rate
convergence and this type of partition functions can effect their
computation of EC rates at high temperatures. Moreover our pn-QRPA
model incorporates a large model space of 5$\hbar\omega$, which can
efficiently handle all parent and daughter excited levels considered
in this calculation. Another distinguishing feature of the current
calculation is that we do not take into account the Brink-Axel
hypothesis (BAH) in our calculation of EC rates, used by IPM and
LSSM calculations. BAH presumes that the transition strengths from
excited parent levels are same as ground-state distribution, drifted
\textit{only} by the excitation energy of the level. We did a
state-by-state computation for EC rates from all parent to daughter
levels in a microscopic way. It is suggested that core-collapse
simulators may check the effect of our enhanced EC weak-rates. For
the case of $^{55}$Mn (Fig.~6), the LSSM and pn-QRPA weak-rates are
in decent agreement till T$_{9}$ = 10. This is due to the fact that
ground-state rate commands the total EC rates for $^{55}$Mn and both
LSSM and pn-QRPA models performed a microscopic calculation of
ground-state EC rates.

\section{Conclusions}
EC rates on \emph{fp}-shell isotopes are significant component for
modeling the late phases of stellar evolution that end their lives
either as thermonuclear or core-collapse supernovae. For the
estimation of these stellar weak rates, it is necessary to study the
charge-changing GT transition strength distribution in the EC
direction. Stellar models mostly depend on the theoretical
approaches, which should be tested against the available measured
data.

In this work we have determined the charge-changing GT transitions
strength for odd-A medium-heavy nuclei ($^{45}$Sc and $^{55}$Mn) in
$\beta^{+}$ direction. The model independent Ikeda sum rule was
fulfilled in our results. For stellar applications the EC rates over
wide range of astrophysical density (10 -- 10$^{11}$ g/cm$^{3}$) and
temperature (0.01 GK -- 30 GK) were computed employing the deformed
pn-QRPA model. We compared our results both with theoretical
(including shell and other QRPA models) and measured charge-changing
reaction data. Our results are in decent comparison with the
measured data. We also compared our weak reaction rates with the IPM
and LSSM results. It was concluded that our rates are enhanced in
the presupernova era as compared to previous calculations. From
astrophysical point these enhanced electron capture rates may have
crucial impact on the  late phase evolution of high mass stars and
shock waves energetics. It was also concluded that at high stellar
density and low temperature regions the PE weak-rates may be
neglected in simulation codes.

\section*{Acknowledgements}

J.-U. Nabi would like to acknowledge the support of the Higher
Education Commission Pakistan through project
5557/KPK/NRPU/R$\&$D/HEC/2016 and Pakistan Science Foundation
through project PSF-TUBITAK/KP-GIKI (02). J.-U. Nabi also
acknowledges the support by T\"{u}bitak (Turkey) under Program No.
2221-1059B211700192 where part of this project was completed.

\begin{figure}
\includegraphics [width=5.in]{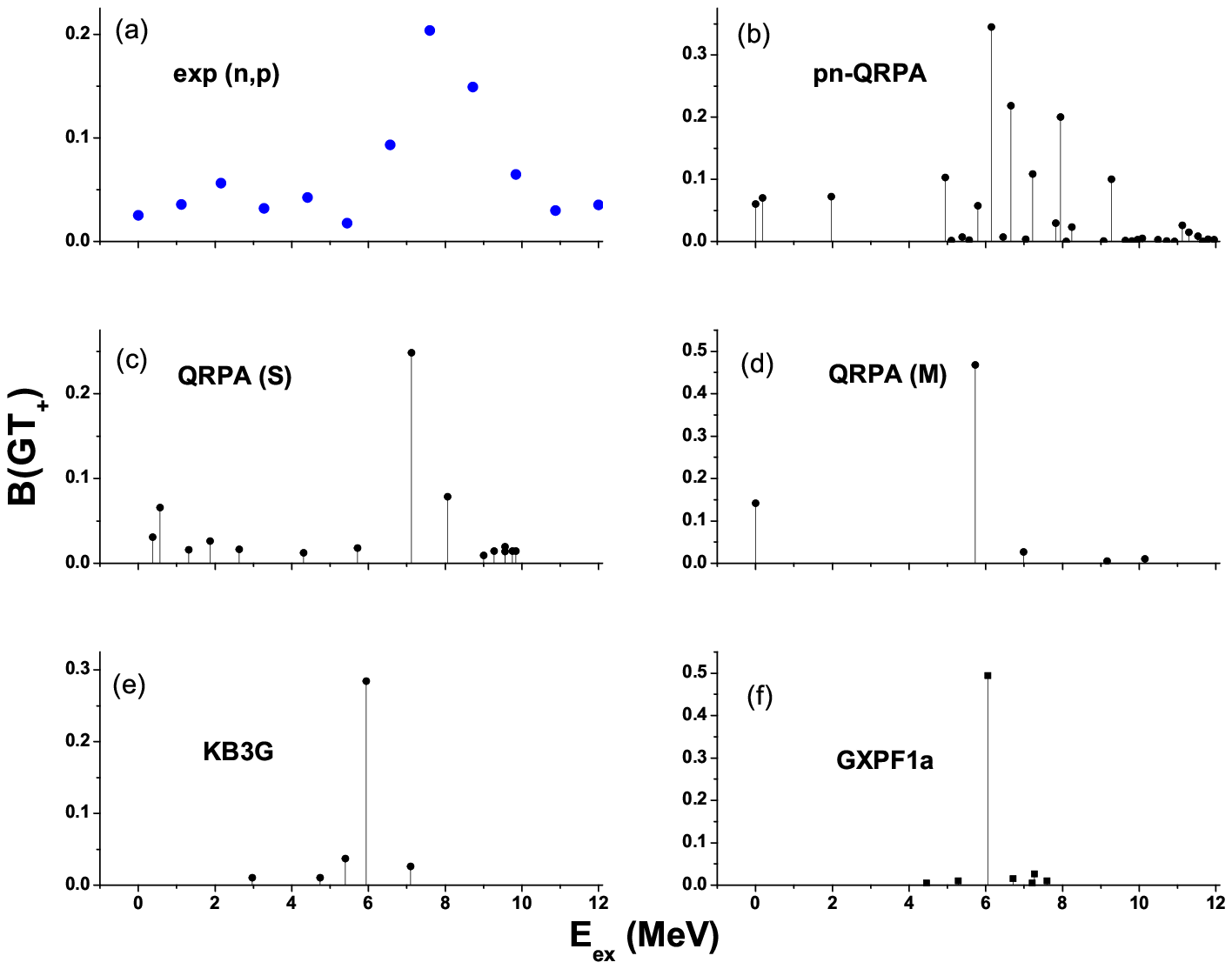}
\centering \caption{ GT transitions from $^{45}$Sc to $^{45}$Ca
plotted versus the excitation energy of the daughter nucleus: (a)
from the (n,p) data of Ref. \cite{Alf91}, (b) from pn-QRPA model
calculations (this work), (c) from the QRPA calculations using the
Skyrme interactions Ref. \cite{Sar13}. Panel~(d) depicts the
theoretical GT transition strength using QRPA formalism of Ref.
\cite{Mol90}. Panels~(e) and (f) show the shell-model calculations
using the KB3G \cite{Pov01} and GXPF1a \cite{Hon04}, interactions
respectively. }\label{fig1}
\end{figure}

\begin{figure}
\includegraphics [width=5.in]{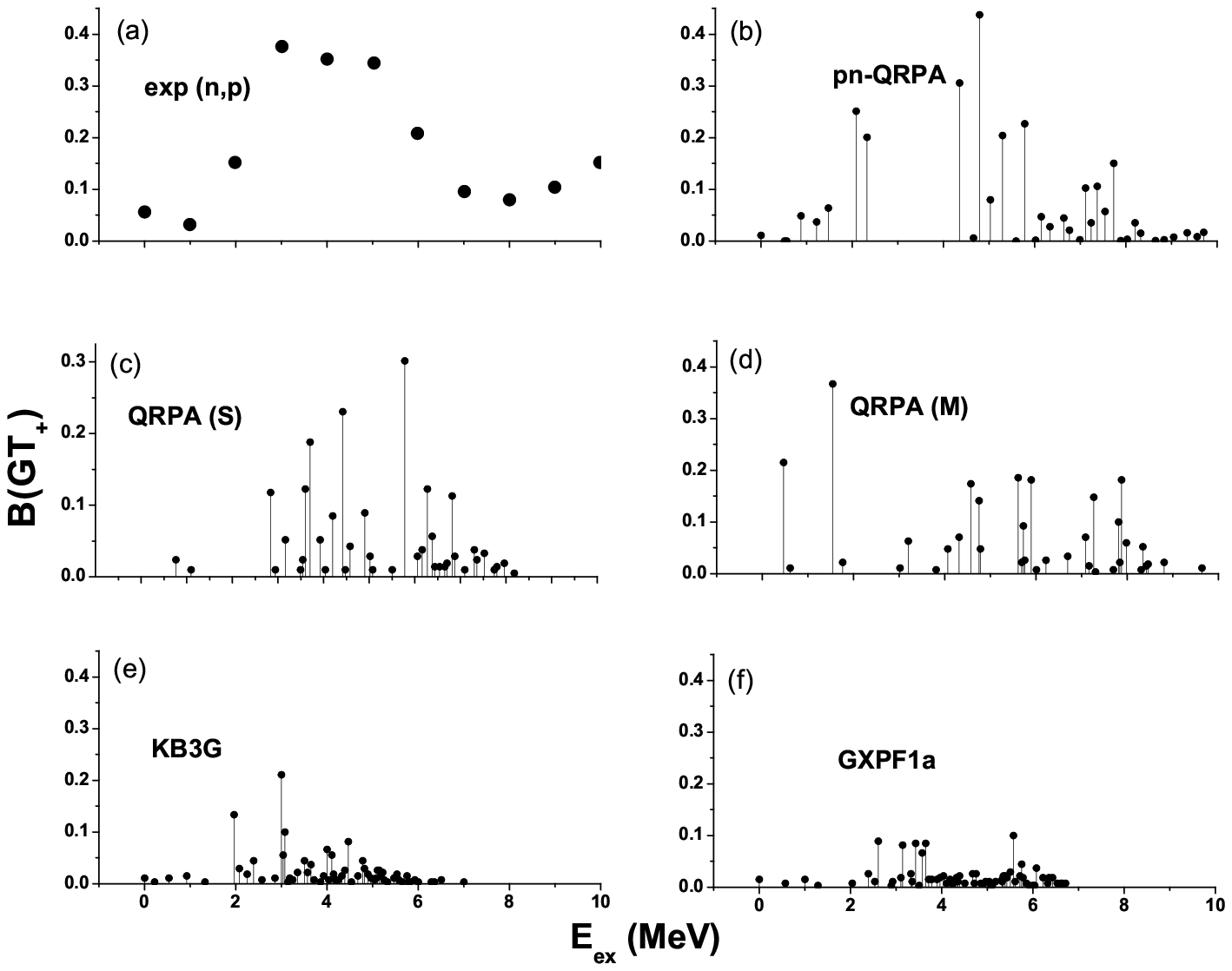}
\centering \caption{Same as Fig.~1, but for $^{55}$Mn.}\label{fig2}
\end{figure}

\begin{figure}
\includegraphics [width=5.in]{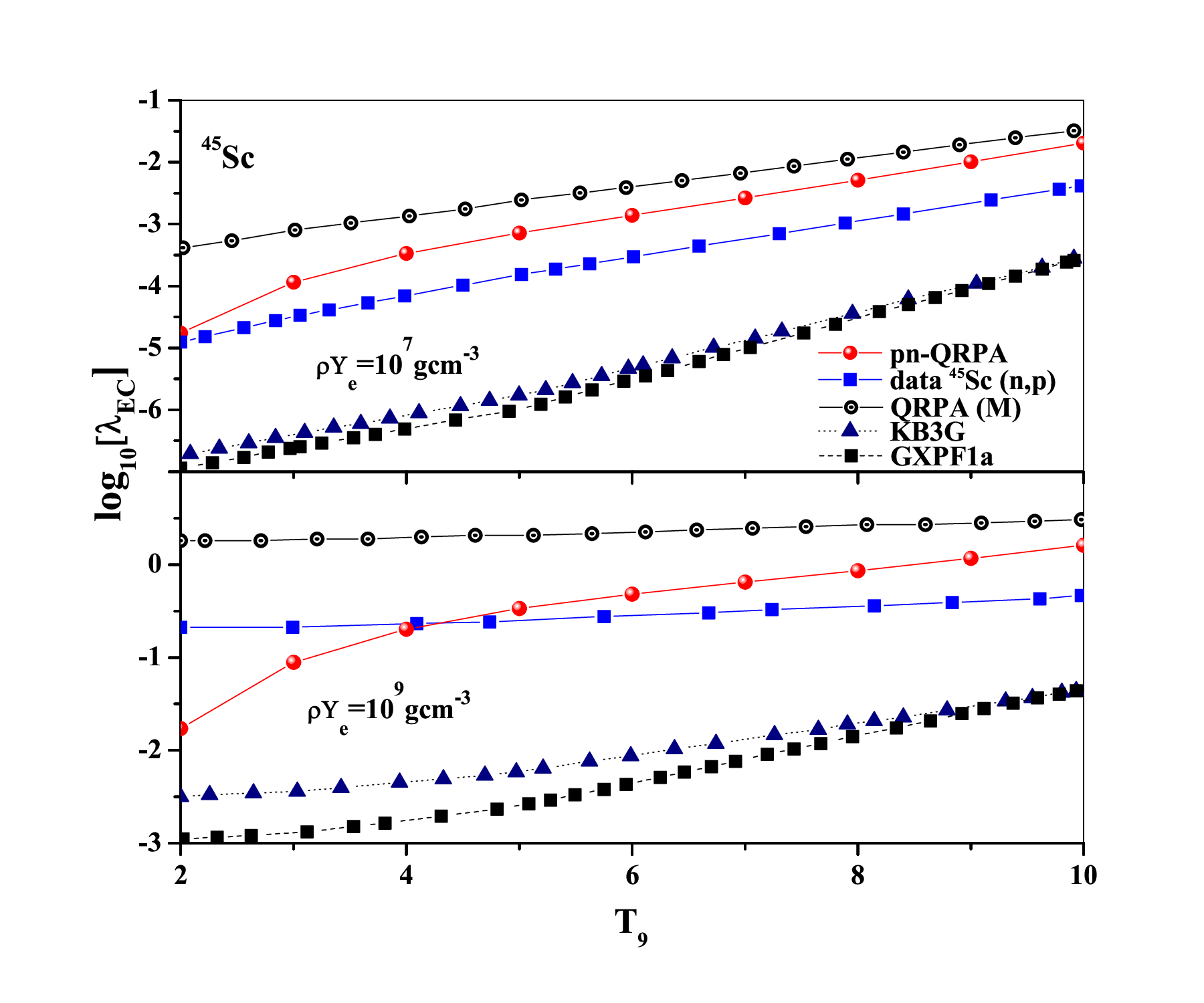}
\centering \caption{Comparison of pn-QRPA calculated electron
capture rates for $^{45}$Sc with data $^{45}$Sc (n,p) \cite{Alf91},
QRPA calculation of Ref. \cite{Mol90} (shown as QRPA (M)), and with
shell-model calculations using the KB3G \cite{Pov01} and GXPF1a
\cite{Hon04}, interactions respectively. $\rho$Y$_{e}$ show the
stellar density, temperature (T$_{9}$) is given in units of 10$^{9}$
K and $\lambda_{EC}$ represents the EC rates in units of s$^{-1}$.
}\label{fig3}
\end{figure}

\begin{figure}
\includegraphics [width=5.in]{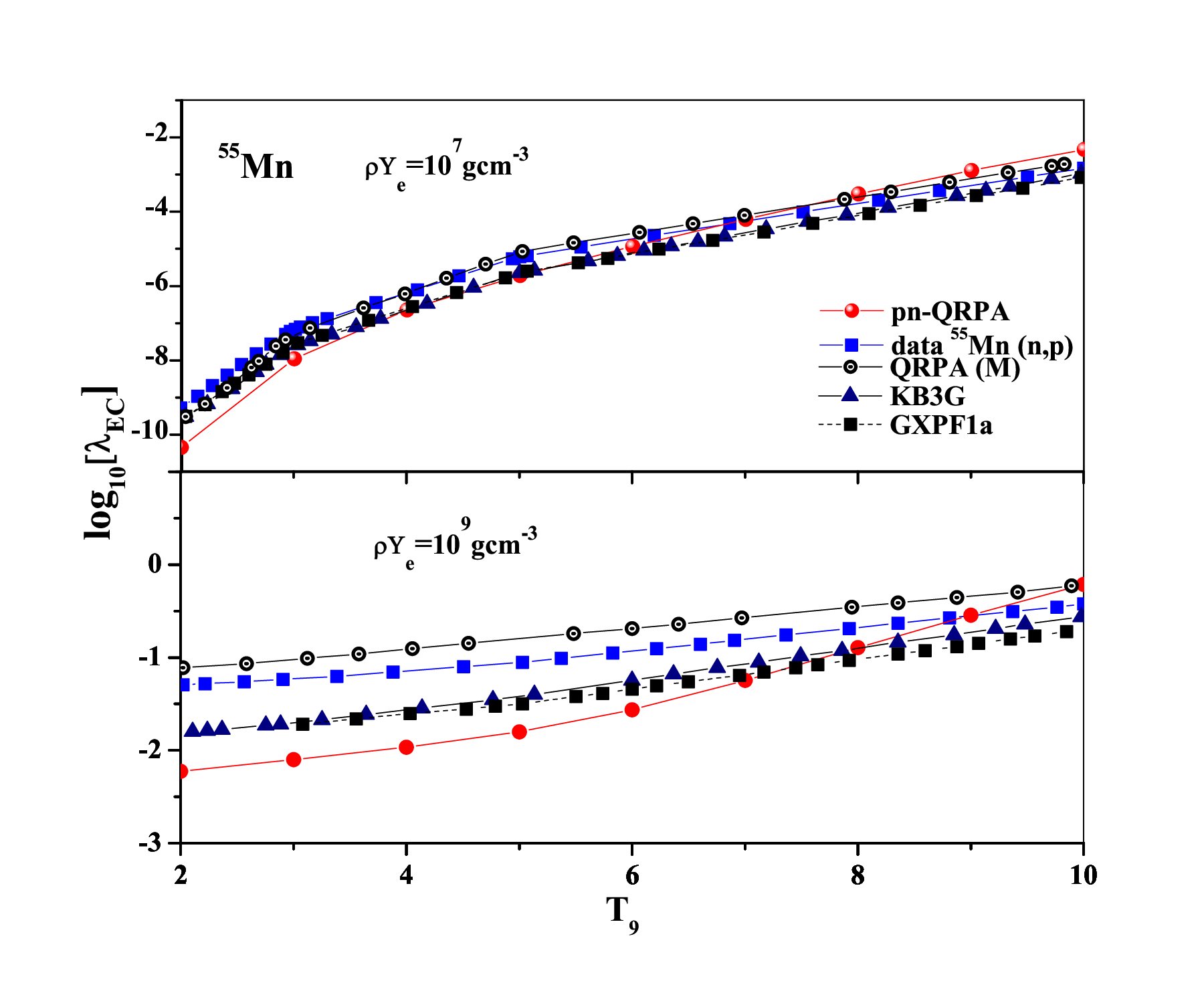}
\centering \caption{Same as Fig~3, but for $^{55}$Mn.}\label{fig4}
\end{figure}

\begin{figure}
\includegraphics [width=5.in]{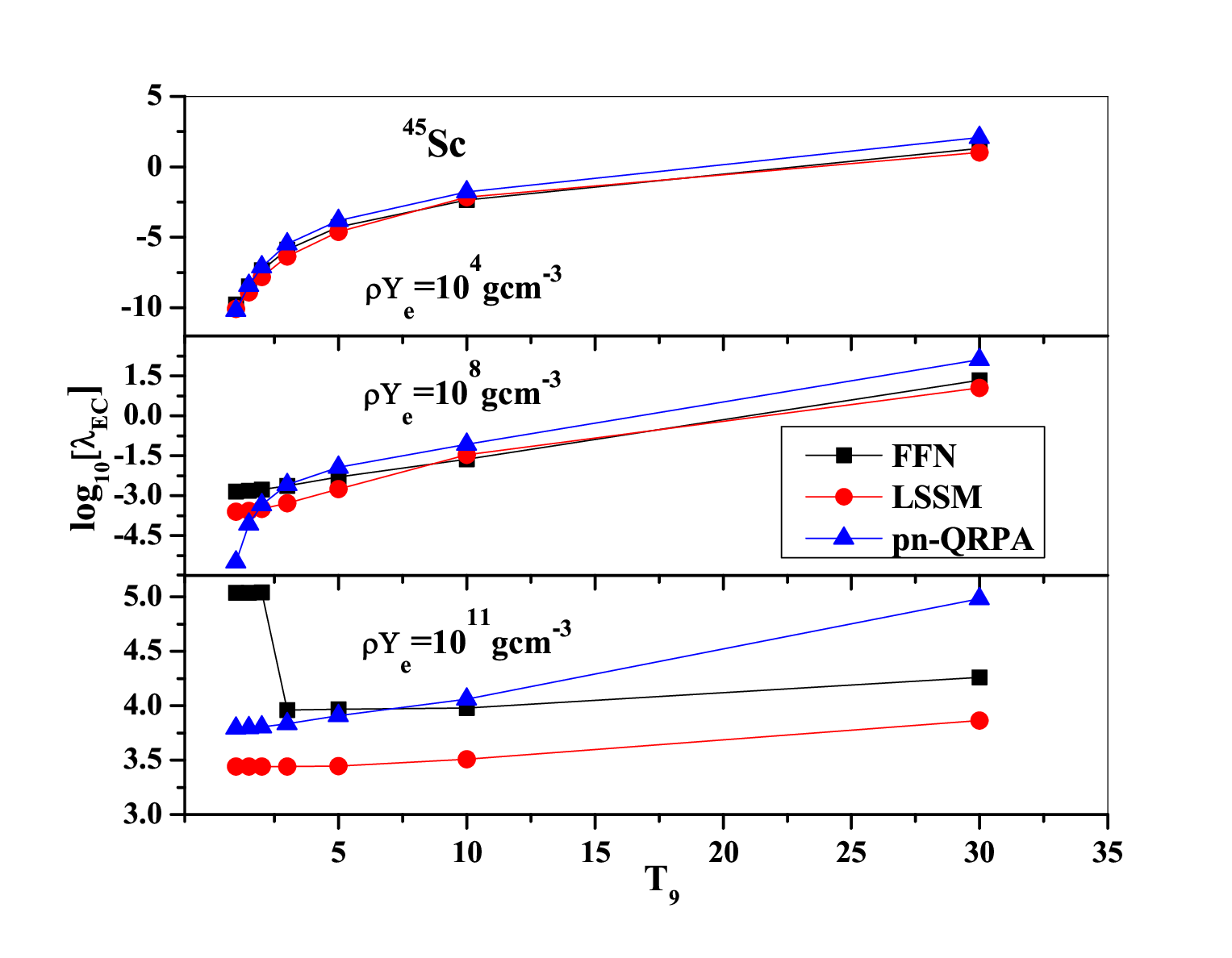}
\centering \caption{Comparison of pn-QRPA calculated electron
capture (EC) rates on $^{45}$Sc, with IPM \cite{Ffn80} and LSSM
\cite{Lan01} calculations. $\rho$Y$_{e}$ show the stellar density,
temperature (T$_{9}$) is given in units of 10$^{9}$ K and
$\lambda_{EC}$ represents the EC rates in units of
s$^{-1}$.}\label{fig5}
\end{figure}

\begin{figure}
\includegraphics [width=5.in]{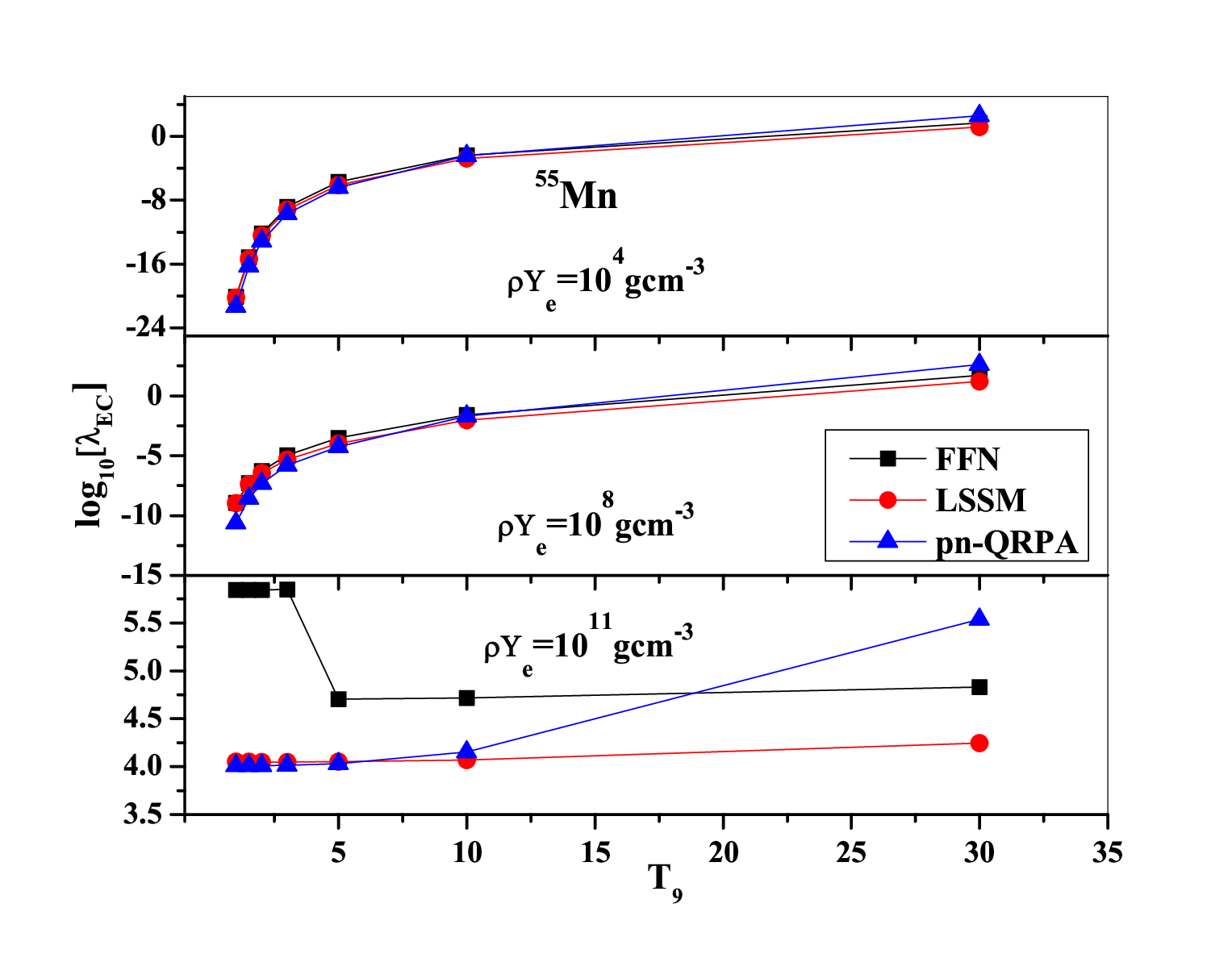}
\centering \caption{Same as Fig~5, but for $^{55}$Mn.}\label{fig6}
\end{figure}

\clearpage
\begin{table}[h]
  \centering
\scriptsize  \caption{Calculated and theoretically predicted values
of Ikeda Sum Rule (re-normalized) for $^{45}$Sc and
$^{55}$Mn.}\label{Table 1}
\begin{tabular}{cccccc}
\hline\hline
A & N & $\sum$B(GT$_{-}$)& $\sum$B(GT$_{+}$)& ISR$_{re}$ (cal) & ISR$_{re}$ (th) \\
\hline
45    & 24    & 4.85 & 1.61  & 3.24 & 3.24 \\
55    & 30    & 8.40 & 3.00  & 5.40 & 5.40 \\
\hline\hline
\end{tabular}
\end{table}

\begin{table}
\scriptsize \caption{Calculated electron capture (EC) and positron
emission (PE) rates in stellar matter for $^{45}$Sc as a function of
stellar density and temperature. $\rho$Y$_{e}$ show the stellar
density (in units of g/cm$^{3}$) and temperature (T$_{9}$) is given
in units of 10$^{9}$ K. The calculated rates are tabulated in log to
base 10 scale and given in units of s$^{-1}$.} \label{Table 2}
\begin{center}
\begin{tabular} {cc|cc|cc|cc}
\hline\hline
 $\rho$$\it Y_{e}$ & T$_{9}$ & EC    & PE & $\rho$$\it Y_{e}$ & T$_{9}$ &  EC    & PE \\
\hline
 10$^{2}$ & 0.7   & -13.999 & -18.130 & 10$^{8}$ & 0.7   & -7.342 & -18.130 \\
 10$^{2}$ & 1     & -11.228 & -14.699 & 10$^{8}$ & 1     & -5.504 & -14.699 \\
 10$^{2}$ & 1.5   & -8.587 & -12.053 & 10$^{8}$ & 1.5   & -4.070 & -12.049 \\
 10$^{2}$ & 2     & -7.130 & -10.754 & 10$^{8}$ & 2     & -3.345 & -10.742 \\
 10$^{2}$ & 3     & -5.472 & -9.491 & 10$^{8}$ & 3     & -2.598 & -9.454 \\
 10$^{2}$ & 5     & -3.808 & -6.785 & 10$^{8}$ & 5     & -1.945 & -6.777 \\
 10$^{2}$ & 10    & -1.788 & -3.322 & 10$^{8}$ & 10    & -1.080 & -3.305 \\
 10$^{2}$ & 15    & -0.308 & -2.198 & 10$^{8}$ & 15    & -0.043 & -2.178 \\
 10$^{2}$ & 20    & 0.746 & -1.681 & 10$^{8}$ & 20    & 0.861 & -1.665 \\
 10$^{2}$ & 25    & 1.502 & -1.402 & 10$^{8}$ & 25    & 1.561 & -1.391 \\
 10$^{2}$ & 30    & 2.075 & -1.236 & 10$^{8}$ & 30    & 2.109 & -1.228 \\
 \hline
 10$^{5}$ & 0.7   & -11.060 & -18.130 & 10$^{11}$ & 0.7   & 3.794 & -18.130 \\
 10$^{5}$ & 1     & -9.207 & -14.699 & 10$^{11}$ & 1     & 3.794 & -14.699 \\
 10$^{5}$ & 1.5   & -7.711 & -12.050 & 10$^{11}$ & 1.5   & 3.796 & -12.049 \\
 10$^{5}$ & 2     & -6.812 & -10.748 & 10$^{11}$ & 2     & 3.802 & -10.742 \\
 10$^{5}$ & 3     & -5.414 & -9.487 & 10$^{11}$ & 3     & 3.832 & -9.454 \\
 10$^{5}$ & 5     & -3.799 & -6.785 & 10$^{11}$ & 5     & 3.907 & -6.777 \\
 10$^{5}$ & 10    & -1.787 & -3.322 & 10$^{11}$ & 10    & 4.060 & -3.300 \\
 10$^{5}$ & 15    & -0.308 & -2.198 & 10$^{11}$ & 15    & 4.319 & -2.152 \\
 10$^{5}$ & 20    & 0.746 & -1.681 & 10$^{11}$ & 20    & 4.600 & -1.609 \\
 10$^{5}$ & 25    & 1.502 & -1.402 & 10$^{11}$ & 25    & 4.819 & -1.307 \\
 10$^{5}$ & 30    & 2.075 & -1.236 & 10$^{11}$ & 30    & 4.981 & -1.121 \\

\hline\hline
\end{tabular}
\end{center}
\end{table}

\begin{table}
\scriptsize \caption{Same as Table~2, but for $^{55}$Mn.}
\label{Table 3}
\begin{center}
\begin{tabular} {cc|cc|cc|cc}
\hline\hline
 $\rho$$\it Y_{e}$ & T$_{9}$ & EC & PE & $\rho$$\it Y_{e}$ & T$_{9}$ &  EC & PE \\
\hline
 10$^{2}$ & 0.7   & -29.155 & -36.954 & 10$^{8}$ & 0.7   & -12.599 & -36.954 \\
 10$^{2}$ & 1     & -22.355 & -28.101 & 10$^{8}$ & 1     & -10.607 & -28.101 \\
 10$^{2}$ & 1.5   & -16.413 & -21.208 & 10$^{8}$ & 1.5   & -8.580 & -21.206 \\
 10$^{2}$ & 2     & -13.186 & -17.668 & 10$^{8}$ & 2     & -7.307 & -17.661 \\
 10$^{2}$ & 3     & -9.684 & -13.436 & 10$^{8}$ & 3     & -5.814 & -13.427 \\
 10$^{2}$ & 5     & -6.438 & -9.331 & 10$^{8}$ & 5     & -4.272 & -9.314 \\
 10$^{2}$ & 10    & -2.440 & -5.911 & 10$^{8}$ & 10    & -1.697 & -5.868 \\
 10$^{2}$ & 15    & -0.280 & -4.798 & 10$^{8}$ & 15    & -0.008 & -4.759 \\
 10$^{2}$ & 20    & 1.034 & -4.285 & 10$^{8}$ & 20    & 1.152 & -4.260 \\
 10$^{2}$ & 25    & 1.922 & -4.010 & 10$^{8}$ & 25    & 1.982 & -3.993 \\
 10$^{2}$ & 30    & 2.571 & -3.847 & 10$^{8}$ & 30    & 2.606 & -3.836 \\
\hline
 10$^{5}$ & 0.7   & -26.135 & -36.954 & 10$^{11}$ & 0.7   & 4.010 & -36.954 \\
 10$^{5}$ & 1     & -20.296 & -28.101 & 10$^{11}$ & 1     & 4.010 & -28.101 \\
 10$^{5}$ & 1.5   & -15.523 & -21.206 & 10$^{11}$ & 1.5   & 4.010 & -21.206 \\
 10$^{5}$ & 2     & -12.863 & -17.665 & 10$^{11}$ & 2     & 4.010 & -17.661 \\
 10$^{5}$ & 3     & -9.624 & -13.434 & 10$^{11}$ & 3     & 4.014 & -13.427 \\
 10$^{5}$ & 5     & -6.428 & -9.331 & 10$^{11}$ & 5     & 4.029 & -9.314 \\
 10$^{5}$ & 10    & -2.439 & -5.911 & 10$^{11}$ & 10    & 4.152 & -5.858 \\
 10$^{5}$ & 15    & -0.279 & -4.798 & 10$^{11}$ & 15    & 4.579 & -4.707 \\
 10$^{5}$ & 20    & 1.035 & -4.285 & 10$^{11}$ & 20    & 5.021 & -4.163 \\
 10$^{5}$ & 25    & 1.922 & -4.010 & 10$^{11}$ & 25    & 5.328 & -3.862 \\
 10$^{5}$ & 30    & 2.571 & -3.847 & 10$^{11}$ & 30    & 5.538 & -3.681 \\

\hline\hline
\end{tabular}
\end{center}
\end{table}

\end{document}